\documentclass{article}

\usepackage{arxiv}

\usepackage[utf8]{inputenc} % allow utf-8 input
\usepackage[T1]{fontenc}    % use 8-bit T1 fonts
\usepackage{hyperref}       % hyperlinks
\usepackage{url}            % simple URL typesetting
\usepackage{booktabs}       % professional-quality tables
\usepackage{amsfonts}       % blackboard math symbols
\usepackage{nicefrac}       % compact symbols for 1/2, etc.
\usepackage{microtype}      % microtypography
\usepackage{lipsum}
\usepackage{graphicx}
\graphicspath{ {./images/} }
\graphicspath{ {./figure/} }
\usepackage{pdfpages}

\title{The initial charge separation step in oxygenic photosynthesis}

\author{
  Yusuke Yoneda,\textsuperscript{1,2,5}
Eric A. Arsenault,\textsuperscript{1,2,3}
Kaydren Orcutt,\textsuperscript{1,2}
Masakazu Iwai,\textsuperscript{2,4}
Graham R. Fleming,\textsuperscript{1,2,3,*}\\
  \textsuperscript{1}Department of Chemistry, University of California, Berkeley, CA 94720, USA\\
  \textsuperscript{2}Molecular Biophysics and Integrated Bioimaging Division, \\
  Lawrence Berkeley National Laboratory, Berkeley, CA 94720, USA\\
  \textsuperscript{3}Kavli Energy Nanoscience Institute at Berkeley, Berkeley, CA 94720, USA\\
  \textsuperscript{4}Department of Plant and Microbial Biology, University of California, Berkeley, CA 94720, USA\\
  \textsuperscript{5}Present Address: Research Center of Integrative Molecular Systems, \\
  Institute for Molecular Science, National Institute of Natural Sciences, Okazaki, Aichi, 444-8585, Japan\\
  \bigskip*grfleming@lbl.gov
}

\begin{document}
\maketitle
\begin{abstract}
Photosystem II is crucial for life on Earth as it
provides oxygen as a result of photoinduced electron transfer and water
splitting reactions. The excited state dynamics of the photosystem
II-reaction center (PSII-RC) has been a matter of vivid debate because
the absorption spectra of the embedded chromophores significantly
overlap and hence it is extremely difficult to distinguish transients.
Here, we report the two-dimensional electronic-vibrational spectroscopic
study of the PSII-RC. The simultaneous resolution along both the visible
excitation and infrared detection axis is crucial in allowing for the
character of the excitonic states and interplay between them to be
clearly distinguished. In particular, this work demonstrates that the
mixed exciton-charge transfer state, previously proposed to be
responsible for the far-red light operation of photosynthesis, is
characterized by the
Chl\textsubscript{D1}\textsuperscript{+}Phe\textsuperscript{-} radical
pair and can be directly prepared upon photoexcitation. Further, we find
that the initial electron acceptor in the PSII-RC is Phe, rather than
P\textsubscript{D1}, regardless of excitation wavelength.
\end{abstract}

\section{Introduction}
    Photosynthesis, the green engine of life on Earth, produces molecular
oxygen by using the light-driven water-plastoquinone oxidoreductase
enzyme known as photosystem II.\textsuperscript{1--3} The photosystem
II-reaction center (PSII-RC) is one of the smallest photosynthetic
components which can undergo charge separation (CS) and thus is an ideal
model system to investigate the underlying mechanism of the initial
light-energy conversion process of photosynthesis.\textsuperscript{4--6}
The PSII-RC consists of six pigments as central cofactors---two special
pair chlorophylls (P\textsubscript{D1} and P\textsubscript{D2}), two
accessory chlorophylls (Chl\textsubscript{D1} and
Chl\textsubscript{D2}), and two pheophytins (Phe\textsubscript{D1} and
Phe\textsubscript{D2})---arranged in a quasi-symmetric geometry (Figure
1a).\textsuperscript{7,8} These six molecules are generally referred to
as RC pigments. In addition, there are two peripheral antenna Chls which
are denoted as Chlz\textsubscript{D1} and Chlz\textsubscript{D2}.
Despite the similarity of the pigment arrangement in the D1 and D2
branches, electron transfer only proceeds along the D1 pigments. The
specifics of how CS proceeds in the PSII-RC is, however, a matter of
vivid debate. In particular, there remains a long-standing discussion
concerned with whether the initial electron acceptor is
P\textsubscript{D1}\textsuperscript{9,10} or
Phe\textsubscript{D1},\textsuperscript{11--13} i.e. whether the initial
radical pair is
(P\textsubscript{D2}\textsuperscript{+}P\textsubscript{D1}\textsuperscript{-})
or
(Chl\textsubscript{D1}\textsuperscript{+}Phe\textsubscript{D1}\textsuperscript{-}).
The uncertainty here is a consequence of the many closely spaced
excitonic states arising from pigment-pigment interactions in the
PSII-RC such that no observable structure is present even in the
electronic linear absorption spectrum at cryogenic
temperatures.\textsuperscript{14--16}

To this end, the excited state dynamics of the PSII-RC has been the
focus of extensive spectroscopic interest spanning over three decades.
These works have included time-resolved
fluorescence,\textsuperscript{17,18} transient
absorption,\textsuperscript{9,10,13,19--21} optical
photon-echo,\textsuperscript{12} visible pump-mid infrared (IR)
probe,\textsuperscript{11} and two-dimensional electronic spectroscopy
(2DES)\textsuperscript{14,22--24} studies. While electronic
spectroscopies acutely suffer from a lack of spectral resolution in
regards to the PSII-RC, the implementation of mid-IR spectroscopy has
proven to be highly advantageous in addressing issues related to
spectral congestion.\textsuperscript{25--28} In particular, the keto and
ester CO stretching modes of Chl and Phe show unique signatures in the
mid-IR region depending on the local protein environment, electronic
structure, and ionic states.\textsuperscript{11,29--33} Additionally,
the amide I modes of the backbone protein can be used as sensitive
reporters for the electron transfer.\textsuperscript{11,31} These were
notably demonstrated by Groot et al. in a visible pump-mid IR probe
study of the PSII-RC where it was suggested that the initial electron
acceptor was Phe based on its distinguishing vibrational
structure.\textsuperscript{11} However, the spectral resolution along
the detection axis alone was not enough to disentangle the distinct
excitonic contributions and dynamics or definitively assign the initial
electron acceptor.

\begin{figure} % picture
    \centering
    \includegraphics[width=16.51cm, height=8.814cm]{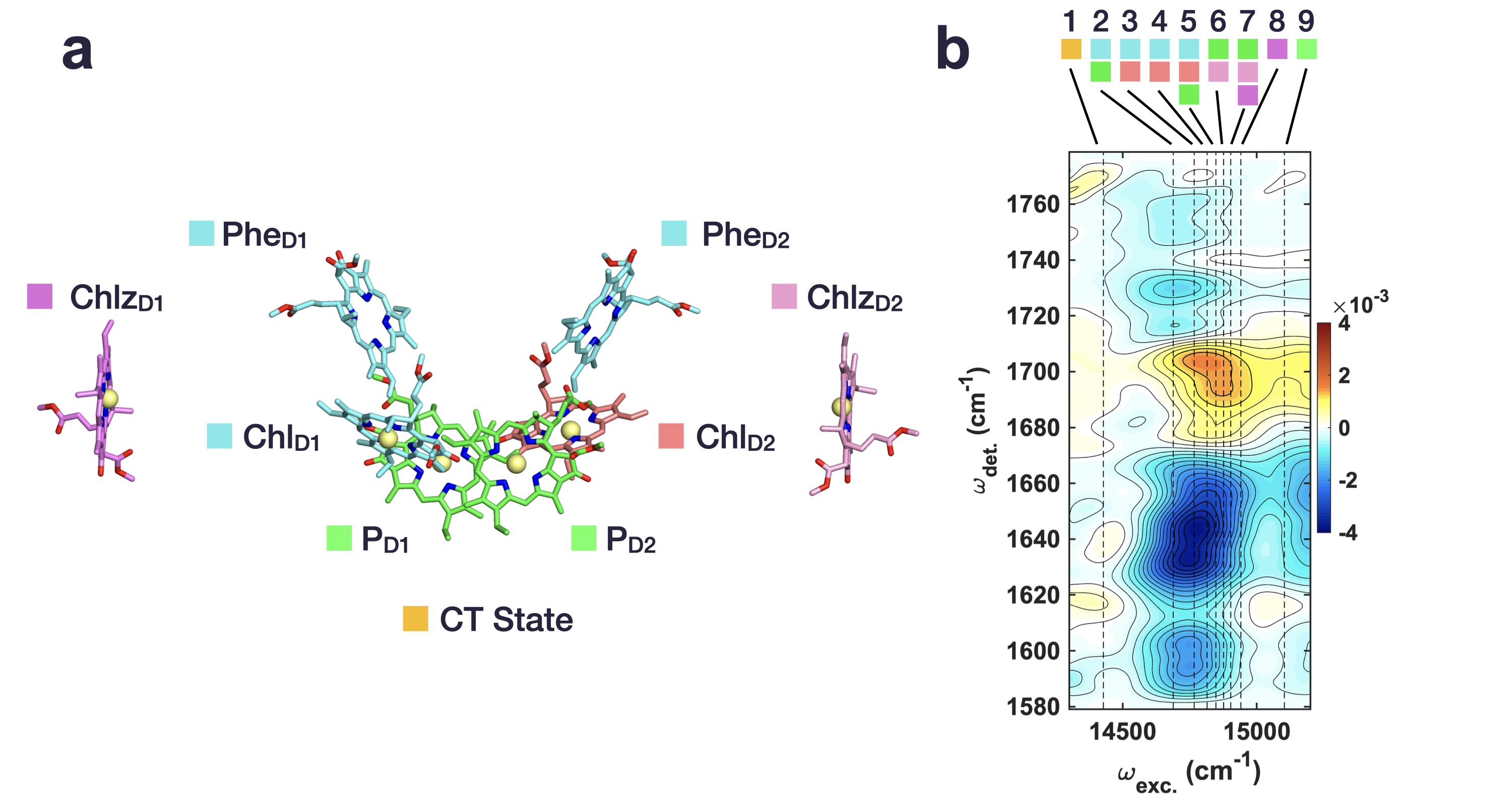}
    \caption{Structure and 2DEV spectrum of the PSII-RC. (a) Pigment arrangement of the PSII-RC depicted based on the crystal
structure (3WU2) reported by Umena et al.\textsuperscript{8} (b) 2DEV
spectrum of the PSII-RC at 170 fs. Positive contours (red/yellow)
indicate ground state bleach (GSB) features and negative contours (blue)
indicate photoinduced absorption (PIA) features. The vertical dotted
lines show the zero phonon exciton transition energies based on the
model by Novoderezhkin et al.\textsuperscript{35} Contour levels are
drawn in 5\% intervals. Colored squares on the top indicate the dominant
pigments participating in each excitonic state as labeled in (a).}
    \label{fig:fig1}
\end{figure}

Many theoretical models have been developed in order to aid in
experimental interpretation and to elucidate the nature the electronic
states at difference absorption wavelengths. Particularly, Stark
spectroscopy suggests that the absorption spectrum of PSII is not
characterized by purely excitonic states, rather it is composed of mixed
exciton-charge transfer (CT) states possibly including contributions
from
(Chl\textsubscript{D1}\textsuperscript{$\delta$+}Phe\textsubscript{D1}\textsuperscript{$\delta$-})*
and
(P\textsubscript{D2}\textsuperscript{$\delta$+}P\textsubscript{D1}\textsuperscript{$\delta$-})*.\textsuperscript{34}
In an attempt to model this, one of the most sophisticated exciton
models of the PSII-RC takes into account eight pigments---the six RC and
two peripheral pigments---and one CS state.\textsuperscript{35} Even in
this model, there was uncertainty as to the character of the initial CS
state because both
P\textsubscript{D2}\textsuperscript{+}P\textsubscript{D1}\textsuperscript{-}
and
Chl\textsubscript{D1}\textsuperscript{+}Phe\textsubscript{D1}\textsuperscript{-}
gave reasonable fits to the data with the former yielding slightly
better agreement to experimental data considered. It is important to
note here that the experimental data was, however, entirely from
electronic spectroscopies.

While uncertainty surrounds the involvement and extent of exciton-CT
mixing in the PSII-RC, studies have suggested that the mixed CT states
are responsible for the far-red excitation of
PSII.\textsuperscript{36--38} Although the absorption of the PSII-RC and
the required redox potential of water oxidation were believed to be
located below 690 nm, it was demonstrated that PSII can be operated by
the far red light beyond 690 nm (exhibiting activities including oxygen
evolution).\textsuperscript{36,39} Additionally, recent EPR
experimental\textsuperscript{37} and QM/MM
theoretical\textsuperscript{38} studies suggest that the far-red light
excitation of PSII involves a lower lying CT state with a hole localized
on Chl\textsubscript{D1} rather than P\textsubscript{D2}. However, just
as spectral congestion obscures the assignment of the initial electron
acceptor, the character of these mixed CT states remains undetermined.

Compared to the previously mentioned techniques, the emerging method of
two-dimensional electronic-vibrational (2DEV) spectroscopy, which
correlates electronic excitation and mid-IR
detection,\textsuperscript{40--44} has the potential to overcome the
challenges associated with congested electronic spectra. In particular,
the simultaneous spectral resolution along both the visible excitation
and IR detection axis has been shown to enable the clear assignment of
transient species.\textsuperscript{41--44} In this study, we
investigated the excited state dynamics of the PSII-RC via 2DEV
spectroscopy. Both highly excitation frequency-dependent spectral
structure and dynamics were clearly resolved. This allowed for a broad
analysis of the excitonic composition of the PSII-RC and direct insight
into the involvement of mixed exciton-CT states found to be directly
prepared upon photoexcitation. Further, the spectra facilitated an
assignment of the initial electron acceptor and enabled the excitation
energy transfer (EET) and electron transfer pathways initiated by
peripheral antenna excitation or RC pigments excitation to be
disentangled.

\section{RESULTS AND DISCUSSION}
\label{sec:headings}
\paragraph{General insights from the 2DEV spectra and IR band assignments.}
    Figure 1b shows the 2DEV spectrum of the PSII-RC 170 fs after
photoexcitation. Of note is the significant excitation frequency
(\emph{$\omega$}\textsubscript{exc.})-dependence of the vibrationally resolved
structure along the detection axis (\emph{$\omega$}\textsubscript{det.}) which,
as we will demonstrate, allows for an excitonic state-specific analysis
of the spectra with high frequency resolution (i.e. vibrationally
resolved excitonic structure). For example, photoinduced absorptions
(PIA) spanning \emph{$\omega$}\textsubscript{det.} = 1,710-1,760
cm\textsuperscript{-1} were seen to clearly favor the lower-lying
excitonic states. Other strong indications of this
\emph{$\omega$}\textsubscript{exc.}-dependent behavior were observed in the
ground state bleach (GSB) region spanning \emph{$\omega$}\textsubscript{det.} =
1,680-1,710 cm\textsuperscript{-1} and the PIAs at
\emph{$\omega$}\textsubscript{det.} = 1,620-1,670 cm\textsuperscript{-1}. These
three regions are of particular interest because, here, vibrational
modes belonging to both the neutral and ionic forms of Chl and Phe can
be clearly distinguished---thus serving as sensitive markers for the EET
and CT steps leading to CS as well as the character of the excitonic
states.

The vibrational structure of the PSII-RC is not only highly
\emph{$\omega$}\textsubscript{exc.}-dependent, but also shows a significant
time-dependence. Therefore, our assignments will be based on the
vibrational structure at specific \emph{$\omega$}\textsubscript{exc.}
corresponding to the energies of exciton 2 (14,690
cm\textsuperscript{-1}) and exciton 8 (14,940 cm\textsuperscript{-1}) in
the model by Novoderezhkin et al.,\textsuperscript{35} which covers the
relevant pigments along the D1 branch, and at either early or later
waiting times (Figure 2).

\begin{figure} % picture
    \centering
    \includegraphics[width=9.474cm, height=10.109cm]{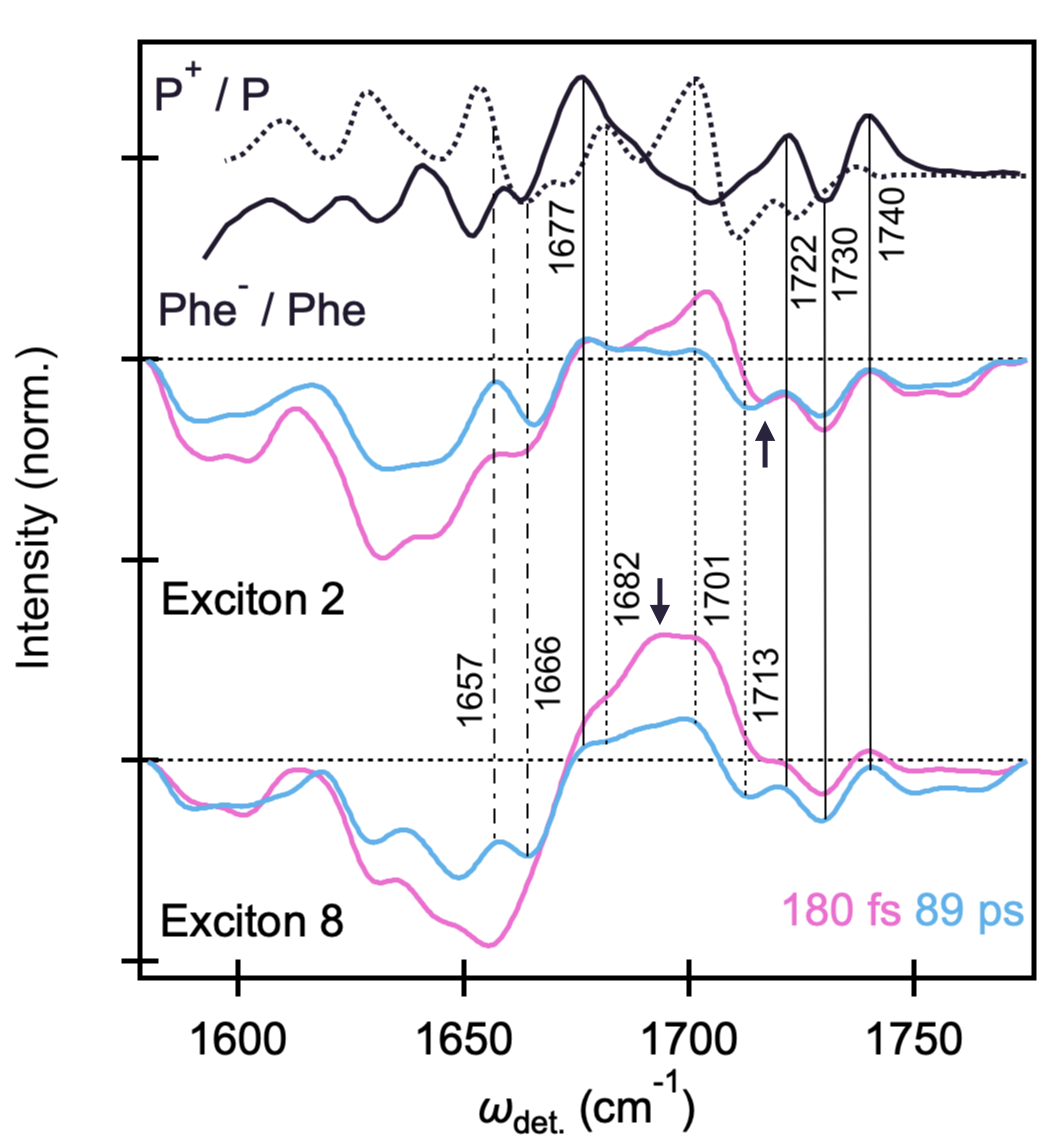}
    \caption{Exciton-specific vibrational
structure and IR assignments. Slices of 2DEV spectrum at
\emph{$\omega$}\textsubscript{exc.} = 14,690 cm\textsuperscript{-1} and
\emph{$\omega$}\textsubscript{exc.} = 14,940 cm\textsuperscript{-1},
corresponding to the energies of exciton 2 and 8 at early (pink, 180 fs)
and later (blue, 89 ps) waiting times. The difference absorption spectra
of P\textsuperscript{+}/P (dotted line) and Phe\textsuperscript{-}/Phe
(solid line) are shown above for comparison (where the signs have been
reversed to match the convention of the 2DEV data). Vertical dotted
(solid) lines indicate to band assignments corresponding
P\textsuperscript{+}/P (Phe\textsuperscript{-}/Phe) while dash-dotted
lines distinguish more ambiguous assignments. The black arrow in exciton
2 marks the Chl\textsubscript{D1}\textsuperscript{+} mode at 1,716
cm\textsuperscript{-1} and in exciton 8 marks the Chlz\textsubscript{D1}
ground state bleach. The P\textsuperscript{+}/P and
Phe\textsuperscript{-}/Phe spectra are reproduced from Ref.
\textsuperscript{30} and \textsuperscript{29} with permission.}
    \label{fig:fig2}
\end{figure}

Generally, the GSB observed at \emph{$\omega$}\textsubscript{det} = 1,680-1,710
cm\textsuperscript{-1} is assigned to the keto CO stretching mode of
Chl/Phe.\textsuperscript{29,31,32} On the electronic ground state, the
frequency of this keto mode depends on the polarity of the environment
and the presence of hydrogen bonding from surrounding media (the larger
the polarity, or the stronger the hydrogen bond, the lower the frequency
of the keto mode). Thus, the GSB can be used to broadly distinguish
pigment contributions (further discussed in the next section). For
example, in Figure 2, it is apparent at early waiting times that the GSB
band of exciton 8 shows much more signal amplitude at 1,680-1,700
cm\textsuperscript{-1} compared to that of the exciton 2. This is in
line with a light-induced FTIR difference spectroscopic study which
reported that Chlz shows a GSB at 1,684
cm\textsuperscript{-1},\textsuperscript{31} whereas P and Phe exhibit
higher and lower frequency GSBs at 1,704 cm\textsuperscript{-1} and 1677
cm\textsuperscript{-1}, respectively.\textsuperscript{29,31,32}

On the electronically excited state, the keto modes of Chl and Phe
exhibit redshifted absorption.\textsuperscript{11,45} For example, in
THF, the keto stretching mode in the previously measured Chl*/Chl
difference spectrum was seen to shift from 1,695 cm\textsuperscript{-1}
to 1,660 cm\textsuperscript{-1}.\textsuperscript{11} Correspondingly,
the negative signal at \emph{$\omega$}\textsubscript{det} = 1,620-1,670
cm\textsuperscript{-1} in both exciton 2 and 8 is broadly assigned to
the excited state absorption (ESA) of the keto modes of Chl and Phe. At
later waiting times, however, there is a notable evolution in the
vibrational structure of this region (Figure 2). Focusing on exciton 2,
a clear dip at 1,657 cm\textsuperscript{-1} appeared concomitantly with
a new peak emerging at 1,666 cm\textsuperscript{-1}. While both the
P\textsuperscript{+}/P and Phe\textsuperscript{-}/Phe difference spectra
exhibit features in this region at frequencies of 1,653-1,655
cm\textsuperscript{-1} and 1,659
cm\textsuperscript{-1},\textsuperscript{29,31,32} respectively, the
signal for Phe\textsuperscript{-}/Phe agrees more closely with the
observed feature at 1,657 cm\textsuperscript{-1}. Resonance Raman
spectroscopy of PSII-RC shows no signal at 1640-1660
cm\textsuperscript{-1}, thus Groot et al. and Noguchi et al. suggest
that the band at 1657 cm\textsuperscript{-1} is assigned to the amide CO
mode reflecting the CS at the RC, rather than keto stretching mode of
Chl or Phe.\textsuperscript{11,31} The band at 1,666
cm\textsuperscript{-1} is similar to both Phe\textsuperscript{-}/Phe and
P\textsuperscript{+}/P showing signal at 1,662 cm\textsuperscript{-1}
and 1,663 cm\textsuperscript{-1},\textsuperscript{29,31,32}
respectively, which has been suggested as a ﻿counterpart of the
previously mentioned band.\textsuperscript{31} A more definitive
assignment is reserved for later discussion.

This leaves the remaining PIA region spanning 1,710-1,760
cm\textsuperscript{-1}. While the ester modes Chl* and Phe* fall in this
region,\textsuperscript{11} they are known to be very weak and would
unlikely account for the full intensity of the observed features.
Further, assuming that this region is only composed of Chl* and Phe*
ester modes would not account for the significant
\emph{$\omega$}\textsubscript{exc.}-dependence clearly present in Figure 1b. If
this was the case, then this region should have a near uniform intensity
across excitons 3 through 7 which have similar pigment contributions and
exciton transition dipole strengths,\textsuperscript{35} but this is
clearly not so (Figure 1b). As a result, contributions from Chl* and
Phe* ester modes are likely small, which should leave this a relatively
clear spectral window, yet, strong features are apparent in the 2DEV
spectra. The Phe\textsuperscript{-}/Phe difference spectrum measured in
PSII, however, shows characteristic signatures in this region, still
related to the ester mode of chromophore itself or surrounding amino
acid residue, with strong absorptions at 1,722 cm\textsuperscript{-1},
1,730 cm\textsuperscript{-1}, and 1,739 cm\textsuperscript{-1} (Figure
2).\textsuperscript{29,32} The corresponding peaks in the 2DEV spectrum
(at 1,722 cm\textsuperscript{-1}, 1,730 cm\textsuperscript{-1}, and
1,740 cm\textsuperscript{-1}), apparent at early waiting times for
exciton 2 and emerging later for exciton 8, are therefore assigned to
Phe\textsuperscript{-}. It should be noted that exciton 8 does show a
slight negative signal around 1,730 cm\textsuperscript{-1} immediately
after photoexcitation, despite being near fully characterized by
Chlz\textsubscript{D1}. We attribute this signal to either slight
contributions from the ester ESA, some degree of overlap between
excitonic bands as these slices only represent the zero phonon
transitions and the actual absorption has finite bandwidth. The ester
mode of the Chl \emph{a} cation (in THF), on the other hand, is known to
blueshift from 1,738 cm\textsuperscript{-1} (neutral) to 1,750
cm\textsuperscript{-1}.\textsuperscript{29} Yet, the
P\textsuperscript{+}/P difference spectrum (Figure 2) does not exhibit
any corresponding characteristic absorptions in this region (the ester
mode of P\textsuperscript{+} appears at 1,743
cm\textsuperscript{-1}).\textsuperscript{30} Thus, the bands in this
region, 1,750 cm\textsuperscript{-1} and 1,764 cm\textsuperscript{-1},
are related to the intermediate Chl cation
(Chl\textsubscript{D1}\textsuperscript{+}) which are also clearly
present in the structure of exciton 2 at early waiting times.

Further characteristic of the Chl \emph{a} cation is a significantly
blueshifted keto stretch, to 1,718 cm\textsuperscript{-1}, (on the order
of 25 cm­\textsuperscript{-1}) versus neutral Chl \emph{a} in
THF.\textsuperscript{33} At early waiting times in exciton 2, for
example, a peak is oberved at 1,716 cm\textsuperscript{-1} which we
assign to Chl\textsubscript{D1}\textsuperscript{+}. However, at later
waiting times, this peak noticeably redshifts to 1,713
cm\textsuperscript{-1}, towards agreement with the characteristic
P\textsuperscript{+} absorption at 1,711 cm\textsuperscript{-1}. This
dynamical behavior will be the focus of later discussion.

To summarize, the significant markers tracking CS in this study are as
follows: Phe\textsuperscript{-} (1,722 cm\textsuperscript{-1}, 1,730
cm\textsuperscript{-1}, and 1,740 cm\textsuperscript{-1}),
Chl\textsubscript{D1}\textsuperscript{+} (at early waiting times: 1,716
cm\textsuperscript{-}­\textsuperscript{1}, 1,750 cm\textsuperscript{-1},
and 1,764 cm\textsuperscript{-1}), and P\textsuperscript{+} (at later
waiting times: 1,713 cm\textsuperscript{-1}). The GSB of the amide CO
bands at 1,657 cm\textsuperscript{-1} and its up-shifted counterpart at
1,666 cm\textsuperscript{-1} reflecting the CS at RC, where the former
likely has predominant contributions from
Phe\textsuperscript{-}\textsubscript{,} while the latter could
potentially be a mixture of Phe\textsuperscript{- ­}and
P\textsuperscript{+}.

\textbf{Excitonic composition and charge transfer character.} Following
the vibrational assignments, we focus on a comparison of the vibrational
structure at specific excitonic energies based on the model by
Novoderezhkin et al.,\textsuperscript{35} in order to understand the
character of the excitonic states and degree of CT mixing. Figure 3a
shows the vibrational structure corresponding to exciton 1, 2, 5, and 8
at an early waiting time. We note again that the exciton energies
discussed thus far are zero phonon lines (shown in Figure 1b). However,
it has been reported that the actual absorption of the CT state shows a
significant blue shift (\textasciitilde5 nm) as a result of coupling to
low-frequency phonons in the environment, compared to other excitonic
bands (1\textasciitilde2 nm).\textsuperscript{35} Thus, to investigate
the CT state specifically, the 2DEV signal corresponding exciton 1 as
shown in Figure 3a was integrated in the range
\emph{$\omega$}\textsubscript{exc} = 14,500-14,650 cm\textsuperscript{-1}.

\begin{figure} % picture
    \centering
    \includegraphics[width=16.51cm, height=8.805cm]{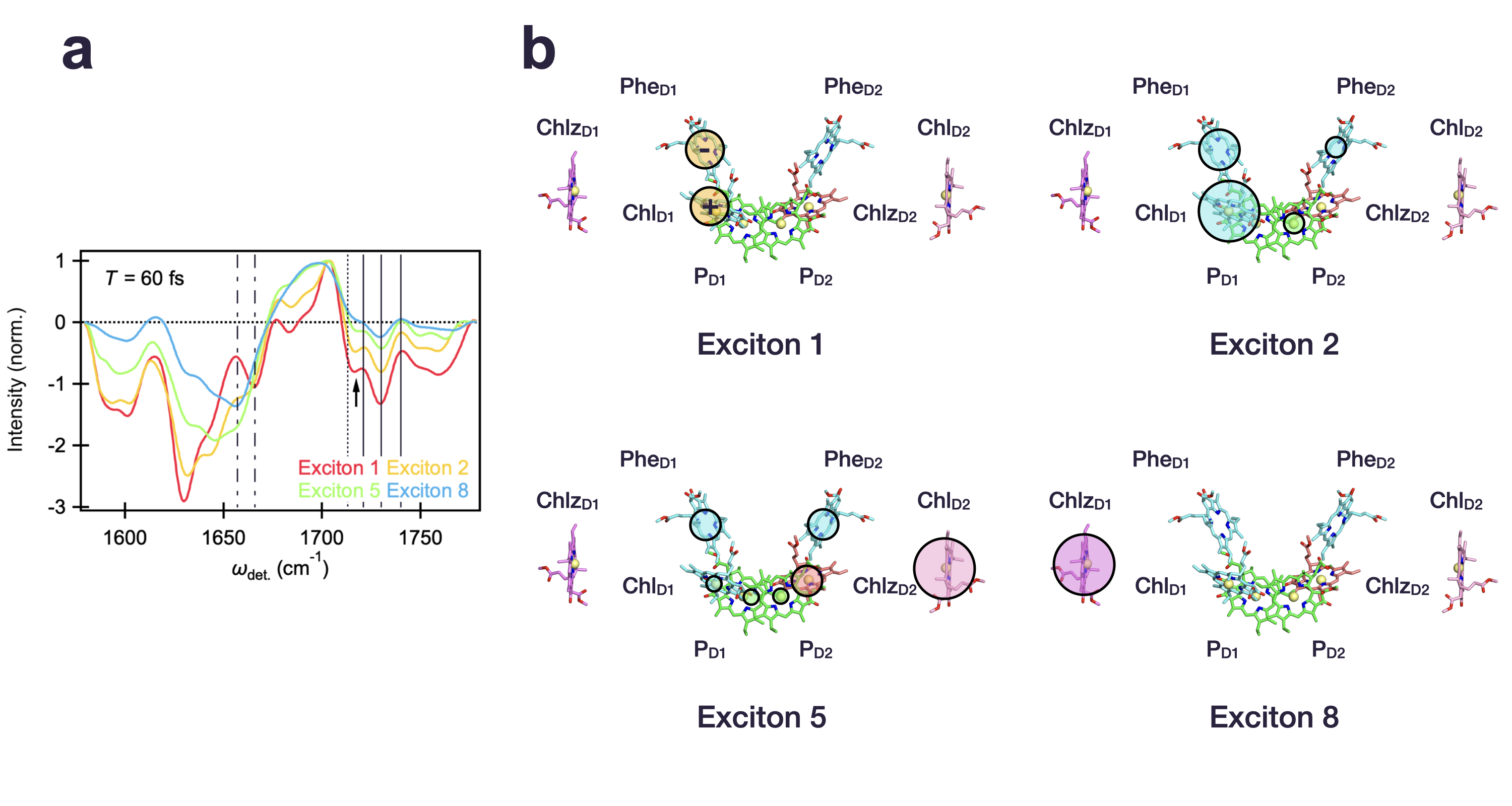}
    \caption{Assignment of excitonic composition and
charge transfer character. (a) Slice along $\omega$\textsubscript{det.} of the
2DEV spectrum corresponding to exciton 1 (red, integrated at
$\omega$\textsubscript{exc.} = 14,500-14,650 cm\textsuperscript{-1}), exciton 2
(yellow, $\omega$\textsubscript{exc.} = 14,690 cm\textsuperscript{-1}), exciton
5 (green, $\omega$\textsubscript{exc.} = 14,850 cm\textsuperscript{-1}), and
exciton 8 (blue, $\omega$\textsubscript{exc.} = 14,940 cm\textsuperscript{-1})
at a waiting time of 60 fs. The vertical solid, dotted, and dash-dotted
lines, as well as the black arrow follow the same convention as in
Figure 2. (b) Character of initial charge transfer state, exciton 1,
along with the site contributions of excitons 2, 5, and 8 where the area
of the shaded circles is proportional to the population of the
corresponding sites based on the model of Novoderezhkin et
al.\textsuperscript{35} For clarity, the slight, additional
contributions from D1 pigments, nearly identical to the relative
contributions of exciton 2, were omitted from exciton 1. Likewise, the
charge transfer character present in excitons 2 and 5 was precluded for
simplicity.}
    \label{fig:fig3}
\end{figure}

At early time, the exciton 1 signal, formed directly upon
photoexcitation, shows clear structure corresponding to
Phe\textsuperscript{-} (1,722 cm\textsuperscript{-1}, 1,730
cm\textsuperscript{-1}, and 1,740 cm\textsuperscript{-1}),
Chl\textsubscript{D1}\textsuperscript{+} (1,716 cm\textsuperscript{-1},
1,750 cm\textsuperscript{-1}, and 1,764 cm\textsuperscript{-1}). In
addition, the amide CO bands reflecting CS at 1,657
cm\textsuperscript{-1} and 1,666 cm\textsuperscript{-1} show clear
structure compared on the other excitonic states, highlighting the
significant CT character of exciton 1 state. The characteristic
P\textsuperscript{+} signal (1,713 cm\textsuperscript{-1}) only appears
at later waiting times and is accompanied by evolution at both of the
aforementioned band positions as well as a decay in the 1,750
cm\textsuperscript{-1} region assigned to
Chl\textsubscript{D1}\textsuperscript{+} (Figure S1)---collectively
indicating a conspicuous lack of initial contributions from
P\textsuperscript{+}.

The lack of P\textsuperscript{+} is in contrast to several previous
spectroscopic studies that suggested there are two CS pathways in the
PSII-RC.\textsuperscript{21,22,24,34} However, these experiments
utilized spectroscopic methods solely in the visible region which are
significantly disadvantaged when it comes to untangling the highly
overlapping signals of the relevant states. In this case, the
vibrational characterization of exciton 1 afforded by the application of
2DEV spectroscopy provides direct evidence that the initial CT state in
the PSII-RC is characterized by
Chl\textsubscript{D1}\textsuperscript{+}Phe\textsuperscript{-} rather
than P\textsubscript{D2}\textsuperscript{+}
P\textsubscript{D1}\textsuperscript{-} (Figure 3b). Such a result is
consistent with a recent QM/MM calculation, utilizing range-separated
TD-DFT theory and the coupled cluster theory with single and double
excitations (CCSD), which proposed that the lowest CT state was
Chl\textsubscript{D1}\textsuperscript{+}Phe\textsuperscript{-}.\textsuperscript{38}
A previous transient IR study also suggested that the initial electron
acceptor is Phe,\textsuperscript{11} however, this study relied on an
extrinsic deconvolution of the vibrational spectrum as opposed to the
intrinsic ability of 2DEV spectroscopy to separate excitonic and CT
contributions along the \emph{$\omega$}\textsubscript{exc.} dimension. This
advantage of 2DEV spectroscopy is particularly useful in the
characterization of the CT state which is only weakly optically allowed
and can therefore be easily obscured in other spectroscopic methods.

Considering the other states, an analysis of the GSB features of exciton
2 and 8 characterize these excitons as predominantly composed of RC
pigments in the active (D1) branch and of the peripheral
Chlz\textsubscript{D1}, respectively, which is consistent with the model
put forth by Novoderezhkin et al. (Figure 3b).\textsuperscript{35} These
assignments also substantiate that Chl and Phe at different binding
position in the PSII-RC are indeed excited by different excitation
frequencies---offering support for the importance of the protein
environment in tuning the site energies of the embedded
pigments.\textsuperscript{38}

Exciton 2 also notably displays characteristic
Chl\textsubscript{D1}\textsuperscript{+} and Phe\textsuperscript{-}
signals at early waiting times (Figure 3a). In comparison to exciton 5,
which is mainly composed of RC pigments in addition to
Chlz\textsubscript{D2} (Figure 3b), these CT signatures in exciton 2 are
markedly more pronounced. Here, we have chosen exciton 5 as a
representative for the energetically intermediate excitonic states,
where there is congestion even in the 2DEV spectra. However, the
vibrational structure is still telling in that the additional
Chlz\textsubscript{D2} contributions of exciton 5 should be similar to
those of Chlz\textsubscript{D1­}, which is indeed reflected in the fact
that exciton 5 resembles a mixture of exciton 2 (mainly RC pigments) and
exciton 8 (mainly composed of a peripheral pigment). This comparison
highlights the enhanced CT character in exciton 2 versus exciton 5 at
early waiting times which confirms the suggestion put forth in the model
by Novoderezhkin et al.\textsuperscript{35} that exciton 2 is
responsible for initiating primary charge separation. Further, in the
model, exciton 1 was taken to be characterized by a CT state which
borrowed intensity from the neighboring state, exciton 2. This is in
agreement with the close resemblance between the GSB and ESA
(particuarly below 1650 cm\textsuperscript{-1} which is outside of the
dominant window for the CS markers) structure of exciton 1 compared to
that of exciton 2 (Figure 3a) and signifies similar overall pigment
contributions. This point is made even clearer on comparison of exciton
1 versus exciton 5 or 8 where there is little similarity in these
regions. Correspondingly, this indicates that exciton 2 is characterized
by a mixed exciton-CT state, rather than a purely excitonic state that
rapidly evolves to the CT state. The mixed character between exciton 1
and 2 also offers a mechanism through which rapid charge separation can
be initiated in the RC.

\textbf{Charge separation dynamics.} To elucidate the dynamics, a global
analysis of the data with sequential modeling was performed. We note
that while the time constants represent a convolution of various
processes, this method is able to holistically capture the spectral
evolution along both frequency dimensions. Therefore, the analysis
captures the \emph{$\omega$}\textsubscript{exc.}-dependent spectra and
dynamics, the latter which can be largely disentangled via vibrational
signatures as we will show. The two-dimensional-evolution associated
difference spectra (2D-EADS) analysis (Figure S1), which can be thought
as the two-dimensional analogue of EADS,\textsuperscript{46} required
five components for a reasonable fit (170 fs, 660 fs, 8.2 ps, 17 ps, and
a non-decaying offset component beyond 100 ps, the duration of the
experiment).

Figure 4 contains exciton-specific slices through the actual 2DEV
spectra along \emph{$\omega$}\textsubscript{det.} at the earliest resolvable
waiting time and at subsequent waiting times corresponding to each of
the above mentioned time constants. Throughout, we focus our attention
on excitons 2, 5, and 8 as these states have substantially more
oscillator strength than exciton 1 and therefore will have a larger
influence on the obtained time constants. The evolution associated with
these time constants can be interpreted such that each spectrum (or
slice) evolves into the next one with the associated time constant. For
example, in exciton 2 (Figure 4b), spectral evolution on the 170 fs
timescale can be understood through a comparison of the pink and yellow
slices. Noticeably, there is growth at 1,657 cm\textsuperscript{-1}, a
characteristic marker for CS. However, in exciton 5 and 8 (Figure 4c and
d, respectively) there is no such growth indicative of CS, rather there
are only slight changes in the keto GSB and ESA regions. On the 660 fs
timescale (comparison of the yellow and green slices in Figure 4b),
exciton 2 exhibits further growth at 1,657 cm\textsuperscript{-1} and
1,666 cm\textsuperscript{-1} while a slight shoulder begins to emerge in
this region for exciton 5. This evolution is also accompanied by marked
changes in the keto ESA structure. We assign both the 170 fs and 660 fs
timescales to progressive completion of CS, i.e.
(Chl\textsubscript{D1}\textsuperscript{$\delta$+}Phe\textsuperscript{$\delta$-})*
\(\longrightarrow\)
Chl\textsubscript{D1}\textsuperscript{+}Phe\textsuperscript{-} (more
pronounced for exciton 2), convoluted with EET within the excitonic
manifold (more pronounced for exciton 5) and an environmental response.
These timescales also agree with previous works which suggested that
there is a fast component to the EET dynamics (100-200 fs time
scale)\textsuperscript{12} and that initial CS occurs within 600-800
fs,\textsuperscript{11} among others which have reported
multiexponential CS dynamics.\textsuperscript{21,24} The distinction
here is that the vibrational structure allows for a targeted assessment
of the dynamical components for each of the states.

\begin{figure} % picture
    \centering
    \includegraphics[width=15.977cm, height=4.369cm]{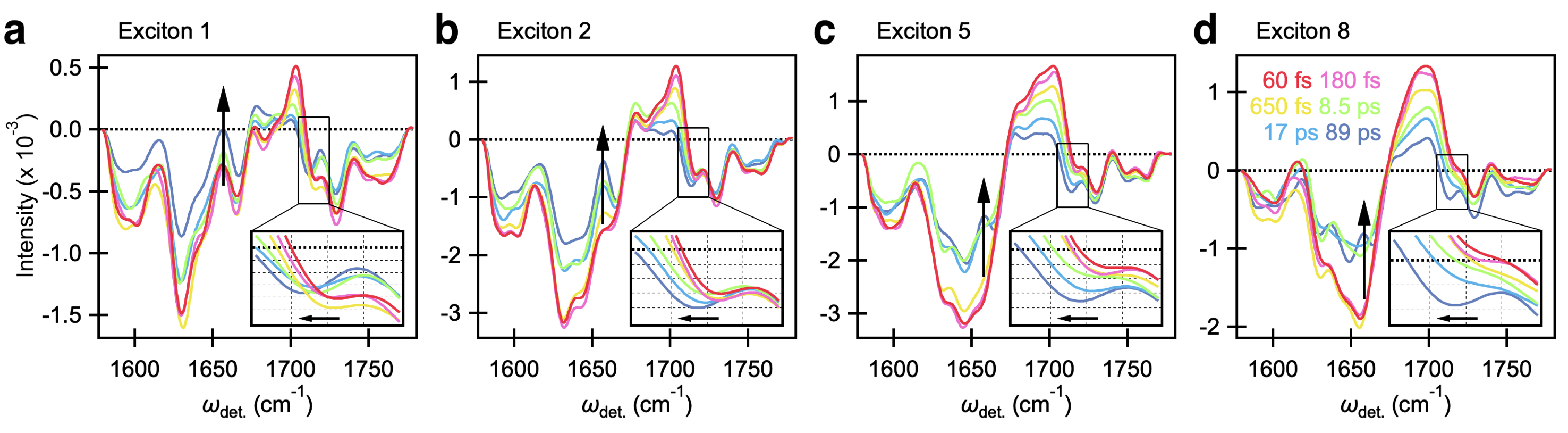}
    \caption{Dynamics of the PSII-RC. The time-dependent
evolution of 2DEV spectra corresponding to excitons 1, 2, 5, and 8.
Inset shows the range of $\omega$\textsubscript{det} = 1705-1725
cm\textsuperscript{-1}, highlighting the red-shifting behavior of the
Chl\textsuperscript{+} band.}
    \label{fig:fig4}
\end{figure}

On an 8.2 ps timescale, both the 1,657 cm\textsuperscript{-1} and 1,666
cm\textsuperscript{-1} CS markers exhibit further evolution along with a
distinct, progressive redshift in the band at 1,716
cm\textsuperscript{-1} to 1,713 cm\textsuperscript{-1} for excitons 1,
2, and 5. This component is similar to the previously reported timescale
for Chl\textsubscript{D1}\textsuperscript{+}Phe\textsuperscript{-}
\(\longrightarrow\) P\textsuperscript{+}Phe\textsuperscript{-} of 6
ps.\textsuperscript{11} Additionally, in a previous light-induced FTIR
difference spectroscopic study, it was proposed that the blue shift of
the keto stretch of Chl cation is smaller for the charge delocalized
dimeric Chl (\textasciitilde10 cm\textsuperscript{-1} in the case of
P680\textsuperscript{+}) compared to that of monomeric Chl
(\textasciitilde30 cm\textsuperscript{-1}).\textsuperscript{47} Both
experimental\textsuperscript{47,48} and
theoretical\textsuperscript{49,50} efforts further support that the P680
cation is partially delocalized over the P\textsubscript{D1} and
P\textsubscript{D2} pigments. Thus, we assign the slight red shift as
the hole migration towards a more delocalized cationic state, i.e.
Chl\textsubscript{D1}\textsuperscript{+}Phe\textsuperscript{-}
\(\longrightarrow\)
(P\textsubscript{D1}P\textsubscript{D2})\textsuperscript{+}Phe\textsuperscript{-}
(likely in addition to further environmental response to CS).
Considering that the mode at 1,713 cm\textsuperscript{-1}, the
characteristic marker for P\textsuperscript{+}, only appears on an 8.2
ps timescale, it is very unlikely that P\textsuperscript{+} contributes
appreciably to the features at 1,657 cm\textsuperscript{-1} and 1,666
cm\textsuperscript{-1} at earlier waiting times. The evolution observed
around 1,657 cm\textsuperscript{-1} and 1,666 cm\textsuperscript{-1} at
later waiting times can therefore be understood as arising from both
Phe\textsuperscript{-} and P\textsuperscript{+}.

The final 17 ps component can be understood as predominantly reflecting
CS limited by EET from peripheral Chlz to RC pigments as only
significant evolution at the CS markers is observed on this timescale
for exciton 8 (Figure 4d). This timescale is also captured by the zero
node line slope (ZNLS) present at \emph{$\omega$}\textsubscript{det} = 1,710
cm\textsuperscript{-1} (Figure 5a, dotted line) in the spectra which
decays with a time constant of 21 ± 4 ps (Figure 5b) and grossly
indicates equilibration within the excitonic manifold. We note that
while the ZNLS trends toward zero, a non-decaying component beyond the
duration of the experiment (\textgreater100 ps) suggests the presence of
the inhomogeneous CS due to the different conformational distributions
of the proteins on the ground state.\textsuperscript{21} This timescale
also falls within the previously established range (14 ps to 37 ps
determined at temperatures of 77 K and 277 K, respectively) for EET from
peripheral Chlz to RC pigments.\textsuperscript{13,19}

\begin{figure} % picture
    \centering
    \includegraphics[width=16.459cm, height=8.128cm]{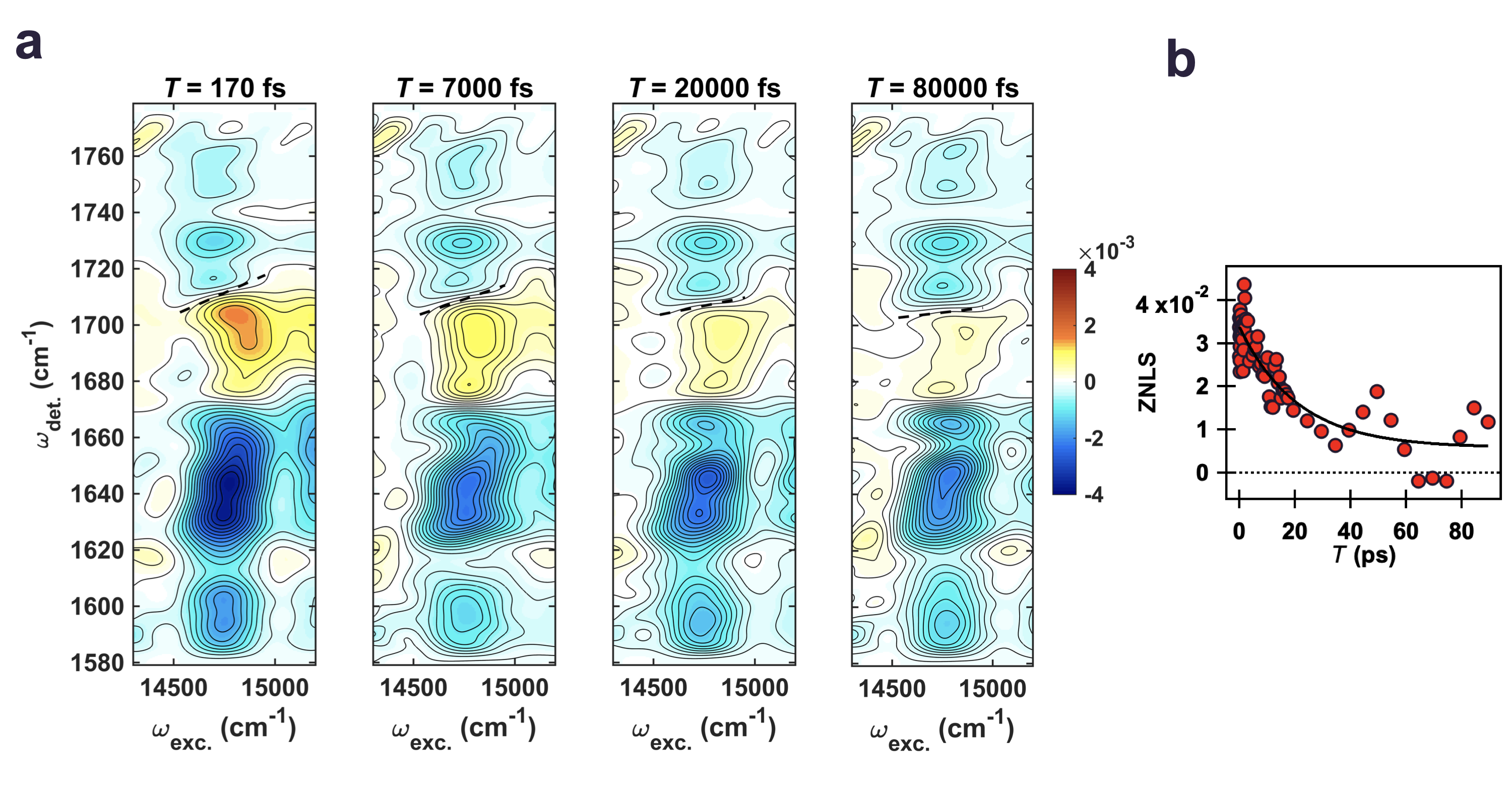}
    \caption{2DEV spectral evolution and ZNLS dynamics of the
PSII-RC. (a) 2DEV spectra of the PSII-RC at different waiting times.
Zero node line slope (ZNLS), obtained by a linear fit of the zero signal
intensity distribution along the excitation axis, is depicted in the
spectra as a dotted line. Contour levels are drawn in 5\% intervals. (b)
ZNLS dynamics of the PSII-RC. Red dots indicate the ZNLS value at each
waiting time and the black curve shows the fit result of a single
exponential function (and an offset) with a time constant of 21 ± 4 ps.}
    \label{fig:fig5}
\end{figure}

\textbf{Concluding comments.} Our results demonstrate that the CT state
can be prepared directly upon photoexcitation, which is characterized by
Chl\textsubscript{D1}\textsuperscript{$\delta$'+}Phe\textsuperscript{$\delta$'-} ($\delta$'
\textgreater{} $\delta$), and indicate that CS is facilitated by exciton-CT
mixing with a contribution from
(Chl\textsubscript{D1}\textsuperscript{$\delta$+}Phe\textsuperscript{$\delta$-})*
throughout the excitonic manifold. The data further establishes that the
initial electron acceptor in the PSII-RC is Phe with no appreciable
competition from P\textsubscript{D1}---independent of excitation
wavelength. These results are entirely in agreement with the recent
theoretical work of Sirohiwal et al. where the
Chl\textsubscript{D1}\textsuperscript{+}Phe\textsuperscript{-} CT state
was found to be the lowest energy excitation globally within the
PSII-RC.\textsuperscript{38} Further, no similarly low energy CT states
involving P\textsubscript{D1}P\textsubscript{D2} were
found,\textsuperscript{38} thus theoretically excluding the special pair
as a candidate for initial CS as our experimental data supports. This is
notably distinct from the bacterial RC where CS is largely initiated at
the special pair (P) with the A branch bacteriochlorophyll (BChl) acting
as the primary acceptor. The distinct excitation asymmetry in the
PSII-RC has been rationalized as a direct consequence of the
electrostatic effect of the protein environment which likely arose as an
evolutionary accommodation of water splitting in oxygenic photosynthetic
systems (particularly its operation in the
far-red).\textsuperscript{36--38} However, this remains an open question
as the initial CS step itself in the has long evaded clear
characterization.

\textbf{ACKNOWLEDGMENTS}

We thank Rafael Picorel for advice regarding isolation of the PSII-RC.
This research was supported by the U.S. Department of Energy, Office of
Science, Basic Energy Sciences, Chemical Sciences, Geosciences, and
Biosciences Division. Y.Y. appreciates the support of the Japan Society
for the Promotion of Science (JSPS) Postdoctoral Fellowship for Research
Abroad. E.A.A. acknowledges the support of the National Science
Foundation Graduate Research Fellowship (Grant No. DGE 1752814).

\textbf{REFERENCES}

1. Wydrzynski, T. J., Satoh, K. \& Freeman, J. A. \emph{Photosystem II
The Light-Driven Water:Plastoquinone Oxidoreductase}. vol. 22 (Springer
Netherlands, 2005).

2. Blankenship, R. E. \emph{Molecular Mechanisms of Photosynthesis, 2nd
Edition}. (Wiley, 2014).

3. Shen, J.-R. The Structure of Photosystem II and the Mechanism of
Water Oxidation in Photosynthesis. \emph{Annu. Rev. Plant Biol.}
\textbf{66}, 23--48 (2015).

4. Renger, G. \& Renger, T. Photosystem II: The machinery of
photosynthetic water splitting. \emph{Photosynth. Res.} \textbf{98},
53--80 (2008).

5. Croce, R. \& Van Amerongen, H. Light-harvesting and structural
organization of Photosystem II: From individual complexes to thylakoid
membrane. \emph{J. Photochem. Photobiol. B Biol.} \textbf{104}, 142--153
(2011).

6. Romero, E., Novoderezhkin, V. I. \& Van Grondelle, R. Quantum design
of photosynthesis for bio-inspired solar-energy conversion.
\emph{Nature} \textbf{543}, 355--365 (2017).

7. Loll, B., Kern, J., Saenger, W., Zouni, A. \& Biesiadka, J. Towards
complete cofactor arrangement in the 3.0 Å resolution structure of
photosystem II. \emph{Nature} \textbf{438}, 1040--1044 (2005).

8. Umena, Y., Kawakami, K., Shen, J.-R. R. \& Kamiya, N. Crystal
structure of oxygen-evolving photosystem II at a resolution of 1.9Å.
\emph{Nature} \textbf{473}, 55--60 (2011).

9. Shelaev, I. V. \emph{et al.} Primary light-energy conversion in
tetrameric chlorophyll structure of photosystem II and bacterial
reaction centers: II. Femto- and picosecond charge separation in PSII
D1/D2/Cyt b559 complex. \emph{Photosynth. Res.} \textbf{98}, 95--103
(2008).

10. Nadtochenko, V. A., Semenov, A. Y. \& Shuvalov, V. A. Formation and
decay of P680 (PD1-PD2) +PheoD1- radical ion pair in photosystem II core
complexes. \emph{Biochim. Biophys. Acta - Bioenerg.} \textbf{1837},
1384--1388 (2014).

11. Groot, M. L. \emph{et al.} Initial electron donor and acceptor in
isolated Photosystem II reaction centers identified with femtosecond
mid-IR spectroscopy. \emph{Proc. Natl. Acad. Sci.} \textbf{102},
13087--13092 (2005).

12. Prokhorenko, V. I. \& Holzwarth, A. R. Primary processes and
structure of the photosystem II reaction center: A photon echo study.
\emph{J. Phys. Chem. B} \textbf{104}, 11563--11578 (2000).

13. Holzwarth, A. R. \emph{et al.} Kinetics and mechanism of electron
transfer in intact photosystem II and in the isolated reaction center:
Pheophytin is the primary electron acceptor. \emph{Proc. Natl. Acad.
Sci.} \textbf{103}, 6895--6900 (2006).

14. Myers, J. A. \emph{et al.} Two-Dimensional Electronic Spectroscopy
of the D1-D2-cyt b559 Photosystem II Reaction Center Complex. \emph{J.
Phys. Chem. Lett.} \textbf{1}, 2774--2780 (2010).

15. Durrant, J. R. \emph{et al.} A multimer model for P680, the primary
electron donor of photosystem II. \emph{Proc. Natl. Acad. Sci. U. S. A.}
\textbf{92}, 4798--4802 (1995).

16. Raszewski, G., Diner, B. A., Schlodder, E. \& Renger, T.
Spectroscopic properties of reaction center pigments in photosystem II
core complexes: Revision of the multimer model. \emph{Biophys. J.}
\textbf{95}, 105--119 (2008).

17. Crystall, B. \emph{et al.} Observation of Multiple Radical Pair
States in Photosystem 2 Reaction Centers. \emph{Biochemistry}
\textbf{30}, 7573--7586 (1991).

18. Konermann, L., Gatzen, G. \& Holzwarth, A. R. Primary processes and
structure of the photosystem II reaction center. 5. Modeling of the
fluorescence kinetics of the D1-D2-cyt-b559 complex at 77 K. \emph{J.
Phys. Chem. B} \textbf{101}, 2933--2944 (1997).

19. Visser, H. M. \emph{et al.} Subpicosecond transient absorption
difference spectroscopy on the reaction center of photosystem II:
Radical pair formation at 77 K. \emph{J. Phys. Chem.} \textbf{99},
15304--15309 (1995).

20. Groot, M. L. \emph{et al.} Charge separation in the reaction center
of photosystem II studied as a function of temperature. \emph{Proc.
Natl. Acad. Sci. U. S. A.} \textbf{94}, 4389--4394 (1997).

21. Romero, E., Van Stokkum, I. H. M., Novoderezhkin, V. I., Dekker, J.
P. \& Van Grondelle, R. Two different charge separation pathways in
photosystem II. \emph{Biochemistry} \textbf{49}, 4300--4307 (2010).

22. Romero, E. \emph{et al.} Quantum coherence in photosynthesis for
efficient solar-energy conversion. \emph{Nat. Phys.} \textbf{10},
676--682 (2014).

23. Fuller, F. D. \emph{et al.} Vibronic coherence in oxygenic
photosynthesis. \emph{Nat. Chem.} \textbf{6}, 706--711 (2014).

24. Duan, H.-G. \emph{et al.} Primary Charge Separation in the
Photosystem II Reaction Center Revealed by a Global Analysis of the
Two-dimensional Electronic Spectra. \emph{Sci. Rep.} \textbf{7}, 12347
(2017).

25. Groot, M. L., Van Wilderen, L. J. G. W. \& Di Donato, M.
Time-resolved methods in biophysics. 5. Femtosecond time-resolved and
dispersed infrared spectroscopy on proteins. \emph{Photochem. Photobiol.
Sci.} \textbf{6}, 501--507 (2007).

26. Di Donato, M. \& Groot, M. L. Ultrafast infrared spectroscopy in
photosynthesis. \emph{Biochim. Biophys. Acta - Bioenerg.} \textbf{1847},
2--11 (2015).

27. Breton, J. Fourier transform infrared spectroscopy of primary
electron donors in type I photosynthetic reaction centers.
\emph{Biochim. Biophys. Acta - Bioenerg.} \textbf{1507}, 180--193
(2001).

28. Noguchi, T. \& Berthomieu, C. Molecular Analysis by Vibrational
Spectroscopy. in \emph{Photosystem II: The Light-Driven
Water:Plastoquinone Oxidoreductase} (eds. Wydrzynski, T. J., Satoh, K.
\& Freeman, J. A.) 367--387 (Springer Netherlands, 2005).
doi:10.1007/1-4020-4254-X\_17.

29. Nabedryk, E. \emph{et al.} Characterization of bonding interactions
of the intermediary electron acceptor in the reaction center of
Photosystem II by FTIR spectroscopy. \emph{Biochim. Biophys. Acta -
Bioenerg.} \textbf{1016}, 49--54 (1990).

30. Breton, J., Hienerwadel, R. \& Nabedryk, E. FTIR Difference Spectrum
of the Photooxidation of the Primary Electron Donor of Photosystem II.
in \emph{Spectroscopy of Biological Molecules: Modern Trends} 101--102
(Springer Netherlands, 1997). doi:10.1007/978-94-011-5622-6\_44.

31. Noguchi, T., Tomo, T. \& Inoue, Y. Fourier transform infrared study
of the cation radical of P680 in the photosystem II reaction center:
Evidence for charge delocalization on the chlorophyll dimer.
\emph{Biochemistry} \textbf{37}, 13614--13625 (1998).

32. Noguchi, T., Tomo, T. \& Kato, C. Triplet formation on a monomeric
chlorophyll in the photosystem II reaction center as studied by
time-resolved infrared spectroscopy. \emph{Biochemistry} \textbf{40},
2176--2185 (2001).

33. Nabedryk, E., Leonhard, M., Mäntele, W. \& Breton, J. Fourier
Transform Infrared Difference Spectroscopy Shows No Evidence for an
Enolization of Chlorophyll a upon Cation Formation either in Vitro or
during P700 Photooxidation. \emph{Biochemistry} \textbf{29}, 3242--3247
(1990).

34. Romero, E. \emph{et al.} Mixed exciton-charge-transfer states in
photosystem II: Stark spectroscopy on site-directed mutants.
\emph{Biophys. J.} \textbf{103}, 185--194 (2012).

35. Novoderezhkin, V. I., Dekker, J. P. \& van Grondelle, R. Mixing of
Exciton and Charge-Transfer States in Photosystem II Reaction Centers:
Modeling of Stark Spectra with Modified Redfield Theory. \emph{Biophys.
J.} \textbf{93}, 1293--1311 (2007).

36. Thapper, A., Mamedov, F., Mokvist, F., Hammarström, L. \& Styring,
S. Defining the far-red limit of photosystem II in Spinach. \emph{Plant
Cell} \textbf{21}, 2391--2401 (2009).

37. Pavlou, A., Jacques, J., Ahmadova, N., Mamedov, F. \& Styring, S.
The wavelength of the incident light determines the primary charge
separation pathway in Photosystem II. \emph{Sci. Rep.} \textbf{8}, 1--11
(2018).

38. Sirohiwal, A., Neese, F. \& Pantazis, D. A. Protein Matrix Control
of Reaction Center Excitation in Photosystem II. \emph{J. Am. Chem.
Soc.} \textbf{142}, 18174--18190 (2020).

39. Pettai, H., Oja, V., Freiberg, A. \& Laisk, A. Photosynthetic
activity of far-red light in green plants. \emph{Biochim. Biophys. Acta
- Bioenerg.} \textbf{1708}, 311--321 (2005).

40. Oliver, T. A. A., Lewis, N. H. C. \& Fleming, G. R. Correlating the
motion of electrons and nuclei with two-dimensional
electronic-vibrational spectroscopy. \emph{Proc. Natl. Acad. Sci.}
\textbf{111}, 10061--10066 (2014).

41. Lewis, N. H. C. \emph{et al.} Observation of Electronic Excitation
Transfer Through Light Harvesting Complex II Using Two-Dimensional
Electronic--Vibrational Spectroscopy. \emph{J. Phys. Chem. Lett.}
\textbf{7}, 4197--4206 (2016).

42. Arsenault, E. A. \emph{et al.} Vibronic mixing enables ultrafast
energy flow in light-harvesting complex II. \emph{Nat. Commun.}
\textbf{11}, 1460 (2020).

43. Arsenault, E. A., Yoneda, Y., Iwai, M., Niyogi, K. K. \& Fleming, G.
R. The role of mixed vibronic Qy-Qx states in green light absorption of
light-harvesting complex II. \emph{Nat. Commun.} \textbf{11}, 6011
(2020).

44. Yoneda, Y. \emph{et al.} Electron--Nuclear Dynamics Accompanying
Proton-Coupled Electron Transfer. \emph{J. Am. Chem. Soc.} \textbf{143},
3104--3112 (2021).

45. Groot, M. L., Breton, J., Van Wilderen, L. J. G. W., Dekker, J. P.
\& Van Grondelle, R. Femtosecond visible/visible and visible/mid-IR
pump-probe study of the photosystem II core antenna complex CP47.
\emph{J. Phys. Chem. B} \textbf{108}, 8001--8006 (2004).

46. Van Stokkum, I. H. M., Larsen, D. S. \& Van Grondelle, R. Global and
target analysis of time-resolved spectra. \emph{Biochim. Biophys. Acta -
Bioenerg.} \textbf{1657}, 82--104 (2004).

47. Okubo, T., Tomo, T., Sugiura, M. \& Noguchi, T. Perturbation of the
structure of P680 and the charge distribution on its radical cation in
isolated reaction center complexes of photosystem II as revealed by
fourier transform infrared spectroscopy. \emph{Biochemistry}
\textbf{46}, 4390--4397 (2007).

48. Diner, B. A. \emph{et al.} Site-directed mutations at D1-His198 and
D2-His197 of photosystem II in Synechocystis PCC 6803: Sites of primary
charge separation and cation and triplet stabilization.
\emph{Biochemistry} \textbf{40}, 9265--9281 (2001).

49. Saito, K. \emph{et al.} Distribution of the cationic state over the
chlorophyll pair of the photosystem II reaction center. \emph{J. Am.
Chem. Soc.} \textbf{133}, 14379--14388 (2011).

50. Narzi, D., Bovi, D., De Gaetano, P. \& Guidoni, L. Dynamics of the
Special Pair of Chlorophylls of Photosystem II. \emph{J. Am. Chem. Soc.}
\textbf{138}, 257--264 (2016).

\textbf{Author contributions}

Y.Y. and G.R.F. conceived the research. Y.Y., E.A.A., and K.O. performed
the 2DEV experiments. Y.Y. analyzed the experimental data. M.I. prepared
the sample. Y.Y., E.A.A., and G.R.F. wrote the manuscript. All authors
discussed the results and contributed to the manuscript.

\textbf{Competing financial interests}

The authors declare no competing financial interests.
 
\includepdf[pages=-]{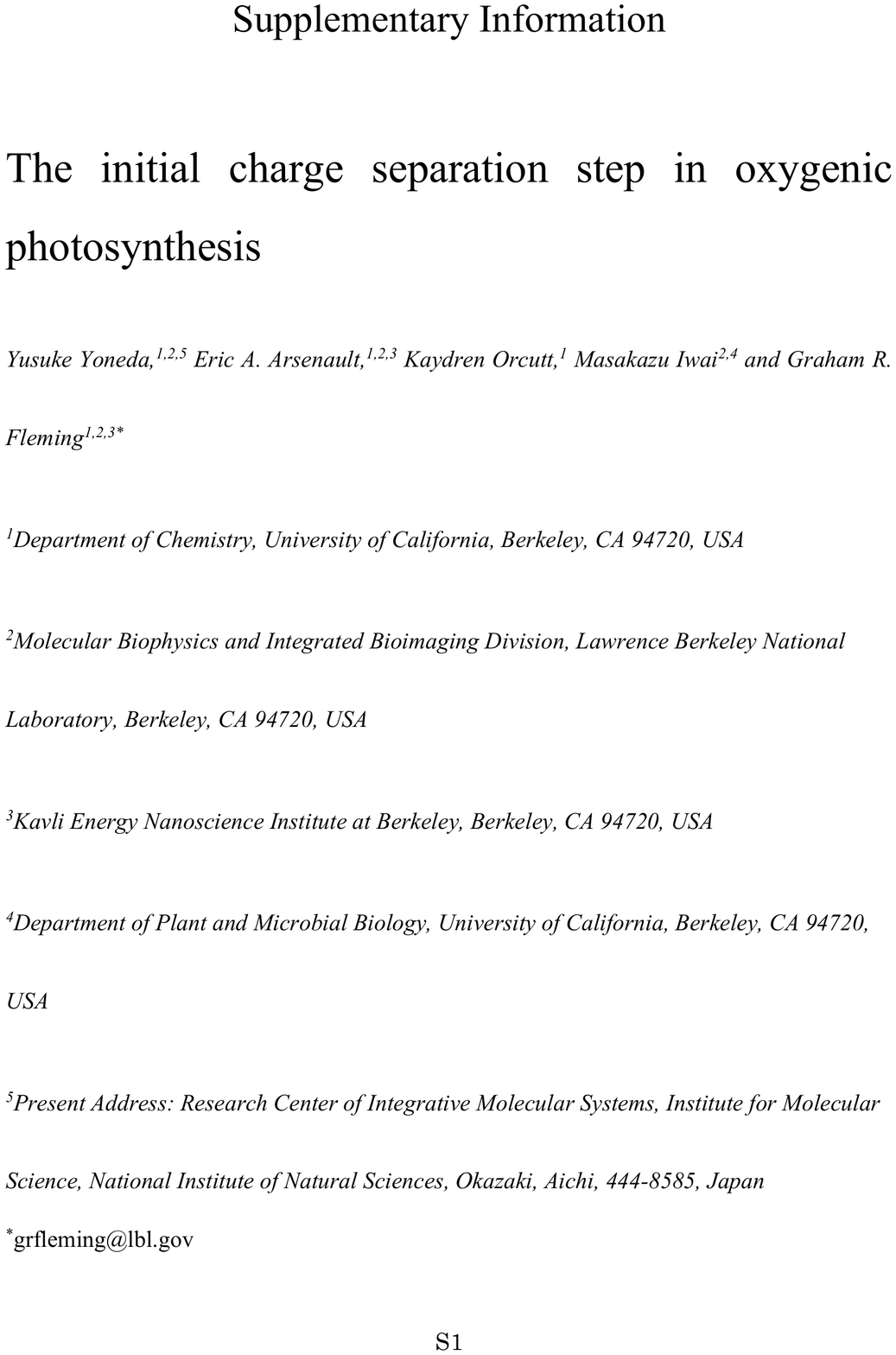}
 
\end{document}